\documentclass[twocolumn,prb,amsmath,amssymb,floatfix,superscriptaddress]{revtex4}
\usepackage{graphicx}
\usepackage{dcolumn}
\usepackage{bm}
\usepackage{color}
\usepackage{array}
\usepackage{float}
\usepackage{supertabular}
\usepackage{longtable}
\begin{document}
\title{Citation Statistics From 110 Years of Physical
  Review\footnote[0]{\copyright 2005 American Institute of Physics. This
    article may be downloaded for personal use only.  Any other use requires
    prior permission of the author and the American Institute of Physics.
    This article appeared in Physics Today, volume 58, page 49, June 2005 and
    the online version may be found at
    http://www.physicstoday.org/vol-58/iss-6/p49.shtml.}}
\author{S.~Redner}\email{redner@cnls.lanl.gov}\altaffiliation{On leave of
  absence from Department of Physics, Boston University.}
\affiliation{Theory Division and Center for Nonlinear Studies, Los Alamos
  National Laboratory, Los Alamos, New Mexico 87545}

\begin{abstract}
  
  Publicly available data reveal long-term systematic features about citation
  statistics and how papers are referenced.  The data also tell fascinating
  citation histories of individual articles.

\end{abstract}

\maketitle

\section*{\large Introduction}

The first particle published in the {\it Physical Review} was received in
1893; the journal's first volume included 6 issues and 24 articles.  In the
20th century, the {\it Physical Review} branched into topical sections and
spawned new journals.  Today, all articles in the {\it Physical Review}
family of journals (PR) are available online and, as a useful byproduct, all
citations in PR articles are electronically available.

The citation data provide a treasure trove of quantitative information.  As
individuals who write scientific papers, most of us are keenly interested in
how often our own work is cited.  As dispassionate observers, we can use the
citation data to identify influential research, new trends in research,
unanticipated connections across fields, and in subfields that are exhausted.
A certain pleasure can also can be gleaned from the data when they reveal the
idiosyncratic features in the citation histories of individual publications.

The investigation of citation statistics has a long history \cite{ER90} in
which a particularly noteworthy contribution was a 1965 study by Derek John
de Solla Price \cite{P65}.  In his study, Price built upon original models by
George Yule and Herbert Simon \cite{Y25} to argue that the distribution in
the number of citations to individual publications had a power-law form.
Price also noted that well-cited papers continue to be referenced more
frequently than less-cited papers, and coined the term cumulative advantage
to describe the mechanism that causes a persistently higher rate\cite{M73}.
Cumulative advantage means that the probability that a publication is cited
is an increasing function of its current number of citations.

In the framework of current fashionable evolving network models, the
mechanism is called preferential attachment \cite{BA99}.  Linear preferential
attachment models provide appealing explanations for the power-law
distributions of connections that are observed in many social systems,
natural networks, and manmade networks such as the World Wide Web \cite{rev}.
One fundamental motivation for studying citation statistics is to determine
whether they exhibit some of the universal features that have been ascribed
to prototypical models of evolving networks \cite{BA99,KRL00,DMS00}.

Before examining the citation data, I offer several caveats: First, the data
include only internal citations --- that is, citations from PR articles to
other PR articles --- and are perforce incomplete.  For highly cited papers,
a previous study \cite{Re04} found that total citations typically outnumber
internal ones by a factor of 3 to 5, a result that gives a sense of the
incompleteness of the PR data.  Second, some 5--10\% of citations appear to
be erroneous \cite{Re04,error}, although the recent practice by PR of
crosschecking references when manuscripts are submitted has significantly
reduced the error rate.  Third, papers can be highly cited for many reasons
--- some substantive and some dubious.  Thus the number of citations is
merely an approximate proxy for scientific quality.

\section*{\large Citation distribution and attachment rate}
\label{sec-cite-data}

The PR citation cover 353,268 papers and 3,110,839 citations from July 1893
through June 2003.  The 329,847 publications with at least 1 citation may be
broken down as follows:

\smallskip
{\small\begin{tabular}{rrlcr}

    ~ &  11 &publications with &$>$& 1000 citations\\
   ~  &  79 &publications with &$>$&  500 citations\\
  ~  &  237 &publications with &$>$&  300 citations\\
 ~  &  2,340 &publications with &$>$&  100 citations\\
 ~  &  8,073 &publications with &$>$&   50 citations\\
 ~ & 245,459 &publications with &$<$&   10 citations\\
 ~ & 178,019 &publications with &$<$&    5 citations\\
 ~ &  84,144 &publications with & &    1 citation\\
\end{tabular}
}

\smallskip
\noindent A somewhat depressing observation is that nearly 70\% of all PR
articles have been cited less than 10 times.  (The average number of
citations is 8.8.)~  Also evident is the small number of 
highly cited publications; table~\ref{tab-top-10} lists the 11
publications with more than 1000 citations.  

Citations have grown rapidly with time, a feature that mirrors the growth of
PR family of journals.  From 1893 until World War II, the number of annual
citations, from PR publications doubled approximately every 5.5 years.  The
number of PR articles published in a given year also doubled every 5.5 years.
Following the publication crash of the war years, the number of articles
published annually doubled approximately every 15 years.

The citation data naturally raise the question, What is the distribution of
citations?  That is, what is the probability $P(k)$ that a paper gets cited
$k$ times?  This question was investigated by Price, who posited the power
law $P(k)\propto k^{-\nu}$, with $\nu$ positive. A power-law form is exciting
for statistical physicists because it implies the absence of a characteristic
scale for citations --- the influence of a publication may range from useless
to earth-shattering.  The absence of such a characteristic scale in turn
implies that citations statistics should exhibit many of the intriguing
features associated with phase transitions, which display critical phenomena
on all length scales.

Somewhat surprisingly, the probability distribution derived from more than 3
million PR citations still has significant statistical fluctuations.  It
proves useful to study the cumulative distribution, $C(k)=\int_k^\infty
P(k')\, dk'$, the probability that a paper is cited at least $k$ times, to
reduce these fluctuations.  

\begin{figure}[ht] 
 \vspace*{0.cm}
 \includegraphics*[width=0.41\textwidth]{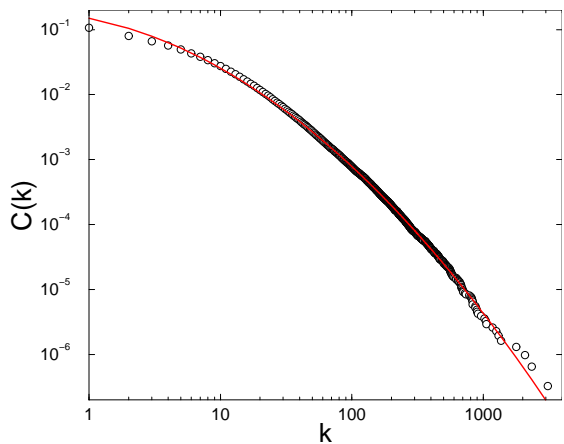}
 \caption{{\bf The cumulative citation distribution} $C(k)$ versus the number
   of citations $k$ for all papers published from July 1893 through June 2003
   in the {\sl Physical Review} journals.  Circles indicate the data.  The
   curve is the log-normal fit $C(k)= A\, e^{-b\ln k - c(\ln k)^2}$, with
   $A=0.15$, $b=0.40$, and $c=0.16$.}
  \label{cumulative}
\end{figure}

On a double logarithmic scale, $C(k)$ has a modest negative curvature
everywhere.  That behavior, illustrated in Fig.~\ref{cumulative}, suggests
that the distribution decays faster than a power law and is at variance with
previous, smaller-scale studies that suggested either a power-law
\cite{P65,R98} or a stretched exponential form \cite{LS98}, $P(k)\propto
\exp(-k^\beta)$, with $\beta<1$.  It is intriguing that a good fit over much
of the range of the range of the distribution is the log-normal form
$C(k)=A\, e^{-b\ln k - c(\ln k)^2}$.  Log-normal forms typically underlie
random multiplicative processes.  The describe, for example, the distribution
of fragment sizes that remain after a rock has been hammered many times.

\begin{figure}[ht] 
 \vspace*{0.cm}
 \includegraphics*[width=0.41\textwidth]{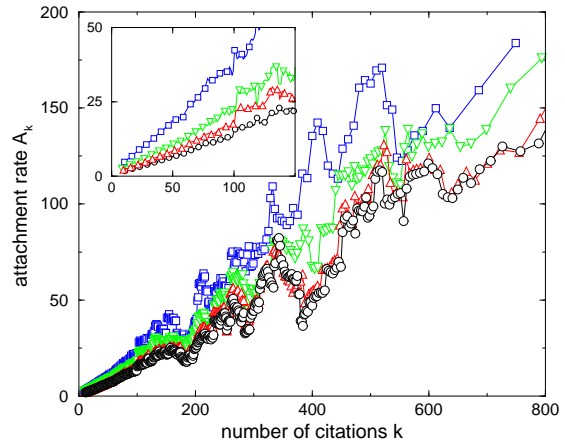}
 \caption{{\bf The attachment rate} $A_k$ is a nearly linear function of the
   number of citations $k$, especially for $k$ less than 150 (inset).  The
   different colors indicate different year ranges for establishing $k$:
   1990-1999 (blue squares), 1980-1999 (green $\bigtriangledown$), 1970-1999
   (red $\bigtriangleup$), and 1893-1999 (black $\circ$).  The rate $A_k$ is
   determined from citations in the year 2000.  Data have been averaged over
   a range of $\pm 2.5\%$.  Other time ranges for existing and new citations
   give similar behavior.
   \label{Ak}}
\end{figure}

The development of citations may be characterized by the attachment rate
$A_k$, which gives the likelihood that a paper with $k$ citations will be
cited by a new article.  To measure the attachment rate, first count the
number of times each paper is cited during a specified time range; this gives
$k$.  Then, to get $A_k$, count the number of times each paper with a given
$k$ in this time window was cited in a subsequent window.  As shown in
Fig.~\ref{Ak}, the data suggest that $A_k$ is a linear function of $k$,
especially for $k< 150$, a condition that applies to nearly all PR papers
\cite{JNB01}.  Thus linear preferential attachment appears to account for the
propagation of citations.

Linear attachment, though, leads to two paradoxes.  First, a linear rate
implies a power-law citation distribution, but Fig.~\ref{cumulative}
indicates that the data are better described by a log-normal form.  While a
log-normal distribution does arise from the nearly linear attachment rate
$A_k=k/(1+a\ln k)$, with $a$ positive, Fig.~\ref{Ak} hints that $A_k$ may be
growing slightly {\em faster} than linearly with $k$.  Second, to implement
linear preferential attachment consciously, a citer must be aware of the
references to every existing paper.  That's clearly not realistic.  A more
realistic process that can lead to linear preferential attachment is the
redirection mechanism \cite{K99,KRL00}.  In redirection, an author who is
writing the reference list for a paper figuratively picks a random paper.
Then the author cites either the random selected paper (with probability
$1-r$) or one of the references within that paper (with probability $r$).
This purely local mechanism generates the linear form $A_k=k+(\frac{1}{r}-2)$
\cite{KRL00}.  Still a mystery is why the myriad of attributes that
influences whether a paper gets cited manifests itself as a citation rate
that is a nearly linear function of the number of citations.

\section*{\large Age structure}  

A common adage says that nobody cites classic papers anymore.  Is this really
true?  How long does a paper continue to get cited?

The age of a citation is the difference between the year when a citation
occurs and the publication year of the cited paper.  Typically, unpopular
papers are cited soon after publication, if at all, and then disappear.  The
converse is also true.  For example, the average citation age $\langle
a\rangle$ over the entire PR data set is 6.2 years, but articles published
before 2000 for which $\langle a\rangle$ is less than 2 years receive, on
average, only 3.6 citations.  On the other hand, highly cited papers usually
continue to be cited for a long time, and vice versa.  Papers with more than
100 citations have $\langle a \rangle= 11.7$ years, and the 11 publications
with more than 1000 citations have $\langle a \rangle=18.9$ years.  For all
PR publications with500 or fewer citations, the average citation age grows
with the number of citations roughly as $\langle a\rangle= k^\alpha$, with
$\alpha\approx 0.3$.

The citation age distributions --- that is, the number of citations a a
function of age --- reveal a fact that is surprising at first sight: The
exponential growth of PR articles strongly skews the form of the age
distributions!  There are, in fact two distinct age distributions. One is the
distribution of {\em citing} ages, defined as the number of citations of a
given age {\em from} a paper.  Remarkably, citing memory is independent of
when a paper is published.  An author publishing now seems to have the same
range of memory as an author who published an article 50 years ago.  The
citing age distribution roughly decays exponentially in age, except for a
sharp decrease in citations during the period of World War II.  However,
citing an old paper is difficult a priori simply because relatively few old
publications exist.  As noted by Hideshiro Nakamoto \cite{N88}, a more
meaningful citing age distribution is obtained by rescaling the distribution
by the total number of publications in each citing year.  So, if one is
interested in journal citations from 2005, the number of cited papers that
are, say, four years old should be scaled by the total number of papers
published in 2001.  The rescaling has a dramatic effect: The nearly
exponential citing age distribution is transformed into a power-law!  An
analogous skewing due to the rapid growth of PR articles also occurs in the
distribution of {\em cited\/} ages, that is, the number of citation of a
given age {\em to} an article.

\section*{\large Individual citation histories}

The citation histories of well-cited publications are diverse from the
collective citation history of all PR articles.  The varied histories roughly
fall into classes that include revived classic works or ``sleeping beauties''
\cite{R04}, major discoveries, and hot publications.  It's fun to examine
examples of each class.

Sometimes a publication will become in vogue after a long dormancy --- a
revival of an old classic.  I arbitrarily define a revived classic as a
nonreview PR articles, published before 1961, that has received more than
$250$ citations and has a ratio of the average citation age to the age of the
paper greater than 0.7.  Thus, revived classics are well-cited old papers
with the bulk of their citations occurring long after publication.  Only the 12
papers listed in table~\ref{tab-classic} fit these criteria.

\begin{figure}[ht] 
 \vspace*{0.cm}
 \includegraphics*[width=0.41\textwidth]{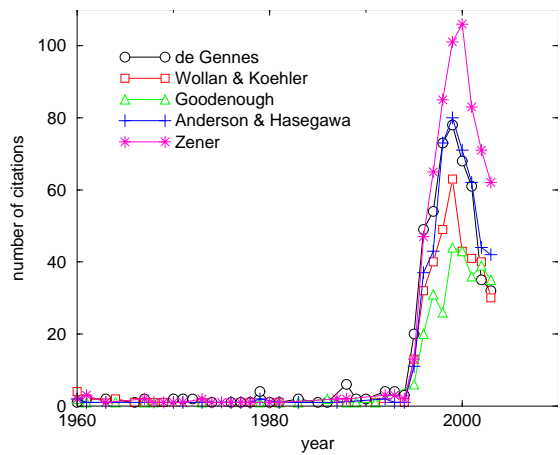}
 \caption{{\bf Five revived classics} of relevance to colossal magnetoresistance
   have similar citation histories.  Table~\protect{\ref{tab-classic}} gives more
   information about these papers.
   \label{rediscovered}}
\end{figure}

The clustered citation histories of the five articles plotted in in
Fig.~\ref{rediscovered} are particularly striking.  These articles, published
between 1951 and 1960 (with three in the same issue of {\it Physical
  Review}), investigated the double exchange mechanism in Perovskite
manganites, the mechanism responsible for the phenomenon of colossal
magnetoresistance.  This topic became in vogue in the 1990's because of the
confluence of new synthesis and measurement techniques in thin-film
transition-metal oxides, the sheer magnitude of the effect, and the clever
use of the term ``colossal''.  The citation burst more than 40 years after
the publication of these five articles is unique in the history of the PR
journals.

The other seven papers have different claims to fame.  Eugene Wigner's 1932
paper had 115 citations before 1980 and 447 through June 2003.  Similarly,
the Albert Einstein, Boris Podolsky and Nathan Rosen (EPR) paper had 36
citations before 1980 and 456 more through June 2003.  With average citation
ages of 55.8 and 59.6 respectively, these are the longest-lived articles with
more than 35 citations in the PR family.  Those papers, as well as the one by
Yakir Aharonov and David Bohm, owe their renewed popularity to the upsurge of
interest in quantum information phenomena.  Wigner's 1934 papers deals with
correlations in an electron gas, a problem of enduring interest in
condensed matter physics.  Julian Schwinger's work is a classic contribution
to quantum electrodynamics.  The 1958 publication by Philip Anderson helped
launched the field of localization.  And Richard Feynman's paper presented a
widely applicable method for calculating molecular forces.  Feynman's article
is noteworthy because it is cited by all PR journals (except the accelerators
and beams special topics journal).

Publications that announce discoveries often receive a citation spike when
the contribution becomes recognized.  I arbitrarily define a discovery paper
has having more than 500 citations and a ratio of average citation age to
publication age less than 0.4; I exclude articles published in {\it Reviews
  of Modern Physics} and compilations by the Particle Data Group.
Table~\ref{tab-discovery} lists the 11 such discovery papers; all were
published between 1962 and 1991.  A trend in this group of papers is the
shift from elementary-particle physics (the six articles published before
1976) to condensed-matter physics (the five articles published after 1983).
The earlier discovery papers reflected major developments in
elementary-particle physics, including $SU(3)$ symmetry, the prediction of
charm, and grand unified theories.  The condensed matter articles are on
quasicrystals, multifractrals, and high-temperature superconductivity.  If
the citation threshold is relaxed to 300, an additional seven papers fit the
discovery criteria.  All of these are concerned with high-temperature
superconductivity, and all but one appear during the golden age of the field,
1987--89.

It is not clear whether the shift in the field of discovery publications
stems from a sea change in research direction or because of from prosaic
concerns.  The past 15 years have seen a major upsurge in quantum
condensed-matter physics that perhaps stems from the discovery of
high-temperature superconductivity.  But recent elementary-particle physics
discoveries may be underrepresented in PR because many CERN-based discoveries
have been published in journals outside the PR family.

\begin{figure}[ht] 
 \vspace*{0.cm}
 \includegraphics*[width=0.41\textwidth]{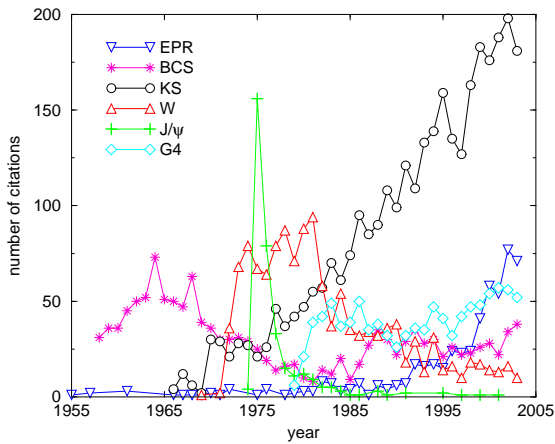}
\caption{{\bf Six classic}, highly-cited publications have varied citation 
  histories.  The abbreviations are defined in the text.
  \label{classic-PT}}
\end{figure}

A number of classic, highly cited publications have noteworthy citations
histories.  Fig.~\ref{classic-PT} illustrates some of these histories.
Citations to ``Theory of Superconductivity'', Phys.\ Rev.\ {\bf 108}, 1175
(1957) by John Bardeen, Leon Cooper and J. Robert Schrieffer (BCS) closely
track the activity in superconductivity; the paper received its fewest
citations in 1985, the year before the discovery of high-temperature
superconductivity.  The BCS paper is the earliest with more than 1000
citations in the PR family.  Steven Weinberg's paper (W) ``A Model of
Leptons'', on the electroweak theory,\ (Phys.\ Rev.\ Lett.\ {\bf 19}, 1264
(1967)), has a broad citation peak followed by a relatively slow decay as
befits this seminal paper's long-term influence.  On the other hand, the
average citation age for the 1974 publications that announced the discovery of
the $J/\psi$ particle -- Phys.\ Rev.\ Lett.\ {\bf 33}, 1404 and 1406 (1974)
-- is less than 3 years!

An unusual example is ``Scaling Theory of Localization: Absence of Quantum
Diffusion in Two Dimensions'', Phys.\ Rev.\ Lett.\ {\bf 42}, 673 (1979) by
Elihu Abrahams, Anderson, Don Licciardello and T. V. Ramakrishnan (the
so-called gang of four; G4).  Since publication, the G4 paper has been cited
30--60 times annually, a striking testament to its long-term impact.  The
paper with the most citations in all PR journals is ``Self-Consistent
Equations Including Exchange and Correlation Effects'', Phys.\ Rev.\ {\bf
  140}, A1133 (1965) by Walter Kohn and Lu Sham (KS).  Amazingly, citations
to this publication have been steadily increasing for nearly 40 years.

The KS paper is also an example of what may be called a hot paper, defined as
a nonreview paper with 350 or more citations, a ratio of average citation age
to publication age greater than two-thirds, and a citation rate increasing
with time.  Ten papers, listed in table~\ref{tab-hot}, fit these criteria.
The 1932 Wigner and 1935 EPR articles, both more than 70 years old, and the
two most-cited PR papers of all time, KS and the 1964 article by Pierre
Hohenberg Kohn, are all still hot.  Astounding!

Of the remaining six hot papers, five are in quantum condensed-matter
physics.  Three of them build on the formalism introduced in the seminal
articles by Hohenberg and Kohn and by Kohn and Sham. Another, Anderson's 1958
localization paper, can be viewed both as hot and as the revival of a
classic.  The newest hot article, by Charles Bennett and coauthors, is
concerned with quantum information theory, a research area that has recently
become fashionable and also led to the sharp increase in citations to
Wigner's 1932 paper and the 1935 EPR paper.

\section*{\large A unique window}

A small number of physicists have played a remarkably large role in top-cited
PR publications.  Two individuals have coauthored five papers from among the
top 100 cited PR articles \cite{Re04}: Kohn, who occupies positions 1, 2, 24,
96, and 100, and Anderson, with positions 9, 19, 20, 35, and 41.  Wigner
appears four times (4, 8, 53, and 55), and Lars Onsager (16, 64, and 68) and
John Slater (12, 27, and 40) each appear three times.

The PR citation data provide a unique window with which to study the
development of citations, and the work I have described can be extended and
applied in many ways.  For example, constructing a graphical representation
of the entire dynamically changing citation network should be revealing.
Such a graph could show how fields develop and could expose unexpected
connections between disparate areas.  A practical, if more controversial, use
of citation data would be to construct retrospective journals that include
only highly cited papers.  Such journals would provide a welcome reduction in
the total literature volume, because only 30\% of all articles have more than
10 citations and a mere 2.3\% have more than 50 citations.  A repository for
all publications would still be necessary, as sleeping beauties do emerge
long after publication.

\acknowledgments{
  
  I thank Mark Doyle of the American Physical Society editorial office for
  providing the citation data, Jon Kleinberg for initial collaboration, Andy
  Cohen and Andy Millis for literature advice, an anonymous referee for
  pointing out Nakamoto's work, Paul Krapivsky and Mark Newman for helpful
  manuscript suggestions, and Claudia Bondila and Guoan Hu for writing Perl
  scripts for some data processing.  Finally, I am grateful to NSF grant
  DMR0227670 (BU) and DOE grant W-7405-ENG-36 (LANL) for financial support.}

\appendix
\begin{widetext}

{\small\begin{longtable}{|p{1.25in}|>{\hfill}p{0.275in}|>{\hfill}p{0.25in}|p{2.8in}|p{2.4in}|}
    \caption{{\sl Physical Review} articles with more than $1000$ citations
      through June 2003.  {\sl PR, Physical Review; PRB, Physical Review B;
        PRD, Physical Review D; PRL, Physical Review Letters; RMP, Reviews of Modern Physics.} }\label{tab-top-10}\\
\hline
              & \#~~    & Av.         &       & \\
  Publication & cites  & Age & Title & Author(s) \\ \hline
\endhead
  PR {\bf 140}, A1133 (1965)&  3227&   26.7& Self-Consistent Equations Including Exchange and Correlation Effects
  & W. Kohn, L. J. Sham \\ \hline
  PR {\bf 136}, B864 (1964)&   2460&   28.7 &  Inhomogeneous Electron Gas&
  P. Hohenberg, W. Kohn \\ \hline
  PRB {\bf 23}, 5048 (1981)&   2079&   14.4&     Self-Interaction Correction
  to Density-Func\-tional Approximations for Many-Electron Systems& J. P. Perdew, A. Zunger\\ \hline
 PRL {\bf 45}, 566 (1980)&   1781&   15.4&      Ground State of the
  Electron  Gas by a Stochastic Method & D. M. Ceperley, B. J. Alder\\ \hline
 PR {\bf 108}, 1175 (1957)&   1364&   20.2 &  Theory of
  Superconductivity&  J. Bardeen, L. N. Cooper, J.~R.~Schrieffer\\ \hline
 PRL {\bf 19}, 1264 (1967)&   1306&   15.5 &   A Model of Leptons&
  S. Weinberg\\ \hline
 PRB {\bf 12}, 3060 (1975)&   1259&   18.4 &   Linear Methods in Band
  Theory&  O. K. Andersen\\ \hline
 PR {\bf 124}, 1866 (1961)&   1178&   28.0 &  Effects of Configuration Interaction on Intensities and Phase Shifts  &  U. Fano\\ \hline
 RMP {\bf 57}, 287 (1985)&   1055&   9.2&     Disordered Electronic
  Systems & P. A. Lee, T. V. Ramakrishnan\\ \hline
 RMP {\bf 54}, 437 (1982)&   1045&   10.8&    Electronic Properties of Two-Dimen\-sional Systems
  & \mbox{T.~Ando, A.~B.~Fowler,} F.~Stern\\ \hline
 PRB {\bf 13}, 5188 (1976)&   1023&   20.8&   Special Points for
  Brillouin-Zone In\-te\-grations&  H. J. Monkhorst, J. D. Pack\\ \hline

\end{longtable}
}

{\small\begin{longtable}{|p{1.15in}|>{\hfill}p{0.25in}|>{\hfill}p{0.25in}|p{3.5in}|p{1.8in}|}
    \caption{The 12 revived classics, as defined in the text, arranged
      chronologically.  {\sl PR, Physical Review.}}\label{tab-classic}\\
    \hline
    & \#~~    & Av.         &       & \\
    Publication & cites & Age & Title & Author(s) \\ \hline \endhead
                                                                                                  
PR {\bf 40}, 749 (1932)&   561&  55.8&        On the Quantum Correction for                                
Thermodynamic Equilibrium & E. Wigner\\ \hline
                                                                                                  
PR {\bf 46}, 1002 (1934)&   557& 51.5& On the Interaction of Electrons in Metals & E. Wigner\\ \hline

PR {\bf 47}, 777 (1935)&   492&  59.6&                                                                     
Can Quantum-Mechanical Description of Physical Reality Be Considered Complete?&                   
\mbox{A. Einstein, B. Podolsky,}\hfil\break ~N.~Rosen\\ \hline
                                      
PR {\bf 56}, 340 (1939)&   342&  49.3& Forces in Molecules & R. P. Feynman\\ \hline  
                                                                                                  
PR {\bf 82}, 403 (1951)&   643&  46.4& Interaction between $d$-Shells in                            
Transition Metals.\ II. Ferromagnetic Compounds of Manganese with Perovskite Structure            
& C. Zener\\ \hline  
                                                                                                  
PR {\bf 82}, 664 (1951)&   663&  36.6& On Gauge Invariance and Vacuum Polarization
& J. Schwinger\\ \hline                                                                       
                                                                                                  
PR {\bf 100}, 545 (1955)&    350&   41.9&                                                                 
Neutron Diffraction Study of the Magnetic Properties of the Series of                             
Perovskite-Type Compounds [$(1-x$)La,$x$Ca]MnO$_3$                                                
&E. O. Wollan, W. C. Koehler \\ \hline
                                                                                                  
PR {\bf 100}, 564 (1955)&   275&  42.0&Theory of the Role of Covalence in                              
the Perovskite-Type Manganites [La, M(II)]MnO$_3$   & J. B. Goodenough  \\ \hline
                                                                                                  
PR {\bf 100}, 675 (1955)&   461&  43.2& Considerations on Double Exchange
& P. W. Anderson, H. Hasegawa\\ \hline 

PR {\bf 109}, 1492 (1958)&   871& 32.0& Absence of Diffusion in Certain Random Lattices &P. W. Anderson\\ \hline

PR {\bf 115}, 485 (1959)&   484&  32.4& Significance of Electromagnetic Potentials in the Quantum Theory
& Y. Aharonov,  D. Bohm\\ \hline 

PR {\bf 118}, 141 (1960)&   500& 37.1&  Effects of Double Exchange in Magnetic Crystals &  P.-G. de Gennes\\ \hline 
\end{longtable}
}

{\small\begin{longtable}{|p{1.2in}|>{\hfill}p{0.25in}|>{\hfill}p{0.25in}|p{3.0in}|p{2.3in}|}  
    \caption{The 11 discovery papers, as defined in the text, arranged
      chronologically.  {\sl PR, Physical Review; PRA, Physical Review
        A; PRB, Physical Review B; PRD, Physical Review D; PRL, Physical
        Review Letters.}}\label{tab-discovery}\\ \hline
          & \#~~    & Av.         &       & \\
  Publication & cites  & Age & Title & Author(s) \\ \hline \endhead
 PR {\bf 125}, 1067 (1962)&   587&  7.0&   Symmetries of Baryons and Mesons & M. Gell-Mann\\ \hline
 PR {\bf 182}, 1190 (1969)&   563& 13.8&    Nucleon-Nucleus
Optical-Model Parameters, $A>40$, $E<50$ MeV   &\mbox{F. D. Becchetti, Jr.,} 
G.~W.~Greenlees\\ \hline 
 PRD {\bf 2}, 1285 (1970)&   738&  11.2&    Weak Interactions with
Lepton-Hadron Symmetry   &\mbox{S. L. Glashow, J. Iliopoulos,} L.~Maiani\\ \hline
 PRL {\bf 32}, 438 (1974)&   545&  11.1&   Unity of All Elementary Forces
& H. Georgi, S. L. Glashow\\ \hline                                                         
PRD {\bf 10}, 2445 (1974)&   577& 11.9&   Confinement of Quarks & 
K. G. Wilson\\ \hline 
 PRD {\bf 12}, 147 (1975)&   501&  10.7&   Hadron Masses in a 
Gauge Theory   &A. De R\'ujula, H. Georgi, S.~L.~Glashow\\ \hline
 PRL {\bf 53}, 1951 (1984)&   559&  7.9&   Metallic Phase with 
Long-Range Orientational Order and No Translational Symmetry&   \mbox{D. Shechtman, I. Blech, 
D. Gratias,} \hfil\break J. W. Cahn\\ \hline                                                 
 PRA {\bf 33}, 1141 (1986)&   501&  6.4&    Fractal Measures and 
Their Singularities: The Characterization of Strange Sets &T. C. Halsey et
al. \\ \hline
 PRL {\bf 58}, 908 (1987)&   625& 1.9&  Superconductivity at 93 K in a \
New Mixed-Phase Yb-Ba-Cu-O Compound System at Ambient Pressure& 
M. K. Wu et al.\\ \hline
 PRL {\bf 58}, 2794 (1987)&   525&  4.8&    Theory of high-$T_c$ Superconductivity in Oxides& 
V. J. Emery\\ \hline
 PRB {\bf 43}, 130 (1991)&   677&  5.2&   Thermal Fluctuations, 
Quenched Disorder, Phase Transitions, and Transport in Type-II Superconductors
&\mbox{D. S. Fisher, M. P. A. Fisher,} D.~A.~Huse\\ \hline  
\end{longtable} 
}

{\small\begin{longtable}{|p{1.25in}|>{\hfill}p{0.29in}|>{\hfill}p{0.25in}|p{2.7in}|p{2.3in}|}
\endhead
\caption{The 10 hot papers, as defined in the text, arranged
  chronologically.  {\sl PR, Physical Review; PRB, Physical Review B; PRL,
Physical Review Letters.}}\label{tab-hot}\\ \hline\endhead
          & \#~~    &Av.& &\\ 
  Publication  & cites & Age&  Title & Author(s) \\ \hline 
PR {\bf 40}, 749 (1932)&   561&   55.8 &     On the Quantum Correction... & E. Wigner\\ \hline
PR {\bf 47}, 777 (1935)&   492&   59.6 &     Can Quantum-Mechanical Description...&  A. Einstein, B. Podolsky, N.~Rosen\\ \hline
  PR {\bf 109}, 1492 (1958)&   871&   32.0&  Absence of Diffusion in Certain Random Lattices &P. W. Anderson\\ \hline
PR {\bf 136}, B864 (1964)&   2460&   28.7 &  Inhomogeneous Electron Gas&  P. Hohenberg, W. Kohn\\ \hline
PR {\bf 140}, A1133 (1965)&  3227& 26.6&     Self-Consistent Equations
Including Exchange...& W. Kohn, L. J. Sham\\ \hline
  PRB {\bf 13}, 5188 (1976)&   1023&   20.8&  Special Points for
  Brillouin-Zone Integrations& H. J. Monkhorst, J. D. Pack\\ \hline
  PRL {\bf 48}, 1425 (1982)&   829&   15.1&  Efficacious Form for Model Pseudopotentials
&L. Kleinman, D. M. Bylander\\ \hline
  PRB {\bf 41}, 7892 (1990)&   691&   9.7&  Soft Self-Consistent Pseudopotentials in a Generalized Eigenvalue Formalism& D. Vanderbilt\\ \hline
  PRB {\bf 45}, 13244 (1992)&   394&   8.1&  Accurate and Simple Analytic Representation of the Electron-Gas Correlation Energy& J. P. Perdew, Y. Wang\\ \hline  
PRL {\bf 70}, 1895 (1993)&   495&   7.4&  Teleporting an Unknown Quantum State via Dual Classical and EPR Channels&
C. H. Bennett, G. Brassard, C. Cr\'epeau, R. Jozsa, A. Peres, W. K. Wootters \\ \hline

\end{longtable}
}

\end{widetext}

\end{document}